\documentclass[
fleqn]{elsart}
\usepackage{amsmath,amssymb}

\def\be{\begin{equation}}
\def\ee{\end{equation}}
\def\bea{\begin{eqnarray}}
\def\eea{\end{eqnarray}}
\def\nn{\nonumber}
\def\p{\partial}

\begin{document}
\begin{frontmatter}
\begin{flushright}
arXiv:0801.0185v3 [hep-th]
\end{flushright}

\title{Covariant anomalies and Hawking radiation from charged rotating black strings in
anti-de Sitter spacetimes}

\author{Jun-Jin Peng},
\author{Shuang-Qing Wu\corauthref{cor}}
\ead{sqwu@phy.ccnu.edu.cn}
\corauth[cor]{Corresponding author.}

\address{College of Physical Science and Technology, Central China Normal University, Wuhan,
Hubei 430079, People's Republic of China}

\begin{abstract}
Motivated by the success of the recently proposed method of anomaly cancellation to derive Hawking
fluxes from black hole horizons of spacetimes in various dimensions, we have further extended the
covariant anomaly cancellation method shortly simplified by Banerjee and Kulkarni to explore the
Hawking radiation of the ($3 + 1$)-dimensional charged rotating black strings and their higher
dimensional extensions in anti-de Sitter spacetimes, whose horizons are not spherical but can be
toroidal, cylindrical or planar, according to their global identifications. It should be emphasized
that our analysis presented here is very general in the sense that the determinant of the reduced
($1 + 1$)-dimensional effective metric from these black strings need not be equal to one $(\sqrt{-g}
\neq 1)$. Our results indicate that the gauge and energy momentum fluxes needed to cancel the ($1 +
1$)-dimensional covariant gauge and gravitational anomalies are compatible with the Hawking fluxes.
Besides, thermodynamics of these black strings are studied in the case of a variable cosmological
constant.
\end{abstract}

\end{frontmatter}


\section{Introduction}

Hawking's discovery \cite{SWH} that a black hole is not completely black but can emit radiation
from its horizon is an interesting and significant quantum effect arising in the background
spacetime with an event horizon. This effect is named as the famous Hawking radiation. It plays
an important role in catching on some clues to the complete theory of Quantum Gravity. This fact
provides a strong motivation for understanding the essence of Hawking radiation since Hawking
discovered it more than thirty years ago. During the past years, one prominent success is that a
lot of new methods to derive Hawking radiation have been developed. For example, Robinson and
Wilczek (RW) \cite{RW} recently presented a novel method to derive the Hawking temperature of a
Schwarzschild black hole via the cancellation of ($1 + 1$)-dimensional consistent gravitational
anomaly. Subsequently, this work was extended to the cases of a charged black hole \cite{IUW} and
rotating charged black holes \cite{RBH} by considering both gauge and gravitational anomalies.
Following this method, a lot of work \cite{OthPs,JWC,KMU,BR,rePs,WP,PW1,BK,WZ,PW2,BKG} appeared
soon. In Refs. \cite{WP}, the original work of \cite{RW,IUW} was generalized to the general case
where the metric determinant $\sqrt{-g} \neq 1$, and a peculiar case in which $\sqrt{-g}$ vanishes
at the horizon has also been investigated \cite{PW1} in details. As far as the case of non-spherical
topology is concerned, only a little work \cite{BR,WP} appeared so far. Therefore it is of special
interest to further investigate Hawking radiation of black objects with non-spherical topology in
other cases.

However, the original anomaly cancellation method proposed in Refs. \cite{RW,IUW} encompasses not
only the consistent anomaly but also the covariant one. Quite recently, Banerjee and Kulkarni \cite{BK}
suggested to simplify this model by considering only the covariant gauge and gravitational anomalies.
Their simplification further cleans the description of the anomaly cancellation method totally in
terms of the covariant expressions, thus making the analysis more economical and conceptually cleaner.
An extension of their work to the case where the determinant $\sqrt{-g} \neq 1$ was done in \cite{WZ}.
Based upon these developments, some direct applications soon appeared in \cite{PW2,BKG}.

On the other hand, with the discovery of anti-de Sitter/conformal-field-theory (AdS/CFT) correspondence
\cite{dSAdSCFT}, it is of great interest to consider rotating charged generalizations of black holes
in AdS spaces. The correspondence between the supergravity in asymptotically AdS spacetimes and CFT
makes it possible to get some insights into the thermodynamic behavior of some strong coupling CFTs
by studying thermodynamics of the asymptotically AdS spacetimes. Specifically speaking, according to
the AdS/CFT correspondence, rotating black holes in AdS spaces are dual to certain CFTs in a rotating
space, while the charged ones are dual to CFTs with a chemical potential.

What is more, higher dimensional generalizations (with or without a cosmological constant) of rotating
black holes and their properties have attracted considerable attention in recent years, in particular,
in the context of string theory, and with the advent of brane-world theories \cite{BW}, raising the
possibility of direct observation of Hawking radiation and of as probes of large spatial extra dimensions
in future high energy colliders. The recent brane-world scenarios that predict the emergence of a TeV-scale
gravity in the higher dimensional theory have opened the door towards testing Hawking effect and exploring
extra dimensions through the decay of mini black holes to high energy particles.

Owing to the above two aspects, therefore it is of great importance to study Hawking radiation of
higher dimensional AdS spacetimes, especially those of non-spherical topology. In this article, we
will employ the covariant anomaly method to investigate Hawking radiation of the ($3 + 1$)-dimensional
charged rotating black strings in AdS spacetimes and their higher dimensional extensions \cite{Awad}.
Our aim is to derive Hawking fluxes from these black strings via covariant gauge and gravitational
anomalies. Our motivation is also due to the fact that the horizon topology of these black strings
is not spherical but can be toroidal, cylindrical or planar, depending on their global identifications.
It should be pointed that our covariant anomaly analysis is very general because the determinant of
the reduced ($1 + 1$)-dimensional effective metric need not be equal to one $(\sqrt{-g} \neq 1)$.
Our results show that the gauge and energy momentum fluxes needed to cancel the ($1 + 1$)-dimensional
covariant gauge and gravitational anomalies are in agreement with the Hawking thermal fluxes. In addition,
thermodynamics of these black strings are studied in the case of a variable cosmological constant.

Our Letter is organized as follows. In Section \ref{fdbs}, we shall investigate thermodynamics of the
($3 + 1$)-dimensional black string and generalize the first law to the case of a variable cosmological
constant. Then we derive the fluxes of the charge and energy momentum tensor via covariant anomalies.
A parallel analysis in higher dimensions is presented in Section \ref{hdbs}, where we have investigated
Hawking radiation from the higher-dimensional black strings. The last section is our summary. An appendix
is supplemented to address the non-uniqueness of two-dimensional effective and physically equivalent
metrics in the process of dimensional reduction, taken the four-dimensional case as an explicit example.

\section{Hawking radiation of the ($3 + 1$)-dimensional
black string via covariant anomalies}\label{fdbs}

We start with the action of Einstein-Maxwell theory in $d$-dimensional spacetimes with a negative
cosmological constant, which reads 
\be
 S_d = -\frac{1}{16\pi}\int d^{d}x\sqrt{-g}\Big[R +\frac{(d-1)(d-2)}{l^2}
  -F^{\mu\nu}F_{\mu\nu}\Big] \, ,
\ee
where $l$ is the radius of AdS spaces. Under this action, a ($3 + 1$)-dimensional rotating charged
black string solution in asymptotically AdS spaces was presented in \cite{Awad}. In terms of
Boyer-Lindquist-like coordinates, its line element and the corresponding gauge potential take
the following forms
\bea
 ds^2 &=& -f(r)\big(\Xi dt -ad\phi\big)^2 +\frac{dr^2}{f(r)} +\frac{r^2}{l^{4}}\big(adt
  -\Xi l^2d\phi\big)^2 +r^2dz^2 \, , \label{bs4d} \\
 A &=& \frac{Q}{r}\big(\Xi dt -ad\phi\big) \, ,
\eea
where
\be
 f(r) = \frac{r^2}{l^2} -\frac{2M}{r} +\frac{Q^2}{r^2} \, , \qquad
 \Xi = \sqrt{1 +\frac{a^2}{l^2}} \, ,
\ee
in which $M$, $Q$, and $a$ are parameters related to the mass, the charge, and the angular momentum,
respectively. When the coordinate $z$ assumes the values $-\infty < z < +\infty$, then the metric
(\ref{bs4d}) describes a stationary black string with a cylindrical horizon. On the other hand,
if the coordinate $z$ is compactified to the region $0 \leq z < 2\pi$, one can get a closed black
string with a toroidal horizon.

We first investigate thermodynamics of the black string. There exists an event horizon at $r_+$,
which is the largest positive root of the equation $f(r) = 0$. The angular velocity and electrostatic
potential of the horizon are
\be
 \Omega = \frac{a}{\Xi l^2} \, , \qquad \Phi = \frac{Q}{\Xi r_+} \, .
\ee
The Hawking temperature $T = \kappa/(2\pi)$ can be obtained via the surface gravity on the horizon,
which is given by
\be
 \kappa = \frac{f^{\prime}(r_+)}{2\Xi} = \frac{3r_+^4 -Q^2l^2}{2\Xi l^2r_+^3} \, .
\ee
For convenience, we assume the length of the black string along the $z$-direction is one. Then the
Bekenstein-Hawking entropy for the horizon is $S = \pi\Xi r^2_H/2$. It can be verified that the mass,
the angular momentum and the electric charge
\be
 \mathcal{M} = \frac{1}{4}\big(3\Xi^2 -1\big)M \, , \qquad
 \mathcal{J} = \frac{3}{4}\Xi Ma \, , \qquad
 \mathcal{Q} = \frac{1}{2}\Xi Q \, ,
\ee
satisfy the integral and differential expressions for the first law \cite{WWXD}
\bea
 && \mathcal{M} = 2TS +2\Omega\mathcal{J} +\Phi\mathcal{Q} +\Theta l \, , \\
 && \delta\mathcal{M} = T\delta S +\Omega\delta\mathcal{J}
  +\Phi\delta\mathcal{Q} +\Theta\delta l \, ,
\eea
where $\Theta = -(2r_+^3 +3Ma^2)/(4l^3)$ is the generalized force conjugate to the variable state
parameter $l$.

We now turn to investigate Hawking radiation from the black string via the covariant anomaly cancellation
method. Since Hawking radiation is the quantum effect due to the horizon, it is very important for us
to analyze the physics near the horizon. To this end, we first show that a free scalar field theory in
the background (\ref{bs4d}) can be reduced to the ($1 + 1$)-dimensional effective theory near the horizon.
For simplicity, we consider the action of a massless scalar field with the minimal electro-magnetic
coupling term
\bea
 S[\varphi] &=& \frac{1}{2}\int d^4x\varphi\mathcal{D}_{\mu}\big(\sqrt{-g}
  g^{\mu\nu} \mathcal{D}_{\nu}\varphi\big) \nn \\
 &=& \frac{1}{2}\int dtdrd\phi dz ~r^2\varphi\Big\{-\frac{\Xi^2r^2 -a^2f(r)}{r^2f(r)}
  \Big[\mathcal{D}_t +\frac{r^2/l^2 -f(r)}{\Xi^2r^2 -a^2f(r)}a\Xi\mathcal{D}_\phi\Big]^2 \nn \\
 &&\quad +\frac{1}{\Xi^2r^2 -a^2f(r)}\mathcal{D}_\phi^2 +\frac{1}{r^2}\p_r \big[r^2f(r)\p_r\big]
  +\frac{1}{r^2}\p_z^2\Big\}\varphi \, , \label{action1}
\eea
where $\mathcal{D}_{\mu} = \p_\mu +ieA_\mu$. After performing a partial wave decomposition $\varphi =
\sum_m\varphi_{m}(t, r)\exp(im\phi +ik_zz)$, and only keeping the dominant terms near the horizon, the
action (\ref{action1}) becomes
\bea
 S[\varphi] &\simeq& \frac{1}{2}\sum_m \int dtdr ~r^2\varphi_m
  \Big\{-\frac{\Xi^2r^2 -a^2f(r)}{r^2f(r)}\Big[\p_t +\frac{ieQ\Xi r}{\Xi^2r^2 -a^2f(r)} \nn \\
&&\quad +ima\Xi\frac{r^2/l^2 -f(r)}{\Xi^2r^2 -a^2f(r)}\Big]^2
  +f(r)\p_r^2\Big\}\varphi_m \, .
\label{effact}
\eea
Therefore, the physics near the horizon can be described by an infinite set of ($1 + 1$)-dimensional
effective massless fields in the background of spacetime with the metric and the gauge potential (see
Appendix for the non-uniqueness of the reduced two-dimensional effective metrics)
\bea
 ds^2 &=& -h(r)dt^2 +\frac{dr^2}{f(r)} \, , \qquad\qquad
  h(r) = \frac{r^2f(r)}{\Xi^2r^2 -a^2f(r)} \, , \label{em1} \\
 \mathcal{A}_t &=& eA_t^{(0)} +mA_t^{(1)} = \frac{eQ\Xi r}{\Xi^2r^2 -a^2f(r)}
  +ma\Xi\frac{r^2/l^2 -f(r)}{\Xi^2r^2 -a^2f(r)} \, ,
\eea
together with the dilaton $\Psi = r\sqrt{\Xi^2r^2 -a^2f(r)}$, which doesn't contribute to the anomalies.
In this effective theory, $A_t^{(1)}$ can be interpreted as an induced $U(1)$ gauge field with the
charge given by the azimuthal quantum number $m$ for each partial wave.

Note that the two-dimensional effective metric (\ref{em1}) is the $r-t$ sector of the four-dimensional
black string spacetime (\ref{bs4d}) in the dragging coordinate system with a dragging velocity $d\phi/dt
= -g_{t\phi}/g_{\phi\phi} = A_t^{(1)}$, and its metric determinant
\be
\sqrt{-g} = \sqrt{h(r)/f(r)} = \frac{r}{\sqrt{\Xi^2r^2 -a^2f(r)}}
\ee
is, in general, not equal to one. It is regular at the horizon ($\sqrt{-g} = 1/\Xi$ when $r = r_+$),
and approaches to one at spatial infinity ($\sqrt{-g} \to 1$ when $r\to \infty$). The latter asymptotic
property is very useful to compute the charge flux and energy flux which are obtained by taking the
$r\to \infty$ limits of the gauge current and energy momentum tensor, respectively. In the remanding
part of this section, we will extend the method in \cite{BK} to analyze Hawking radiation of the black
string through the covariant gauge and gravitational anomalies in the effective background spacetime
(\ref{em1}).

As a warm-up, we briefly review the basic idea of the anomaly cancellation method in which the
quantum fields near the horizon can be effectively described in terms of an infinite collection
of ($1 + 1$)-dimensional massless fields and the underlying effective theory should be invariant.
Specifically speaking, if we omit the classically irrelevant ingoing modes near the horizon, the
($1 + 1$)-dimensional effective field theory becomes chiral, leading to gauge and gravitational
anomalies. Therefore, in order to preserve the invariance of the underlying theory under gauge
and general coordinate transformations, the anomalies must be cancelled by a compensating current,
which is compatible with the Hawking fluxes of charge and energy momentum.

Consider first the charge flux of the black string. It is well known that the covariant gauge
anomaly can be read off as \cite{AWBZBK}
\be
 \nabla_\mu \frac{1}{e}J^{(0)\mu} = \nabla_\mu \frac{1}{m}J^{(1)\mu}
  = \frac{-1}{4\pi\sqrt{-g}} \epsilon^{\alpha\beta}\mathcal{F}_{\alpha\beta} \, ,
 \label{Gpa}
\ee
where $\epsilon^{\alpha\beta}$ is an antisymmetry tensor density with $\epsilon^{tr} = \epsilon_{rt}
= 1$ and $\mathcal{F}_{\alpha\beta} = \p_\alpha\mathcal{A}_\beta -\p_\beta\mathcal{A}_\alpha$.

In our case, there are two $U(1)$ gauge symmetries yielding two gauge currents $J^{(0)r}$ and $J^{(1)r}$,
corresponding to the gauge potentials $A_t^{(0)}$ and $A_t^{(1)}$, respectively. We shall adopt Eq.
(\ref{Gpa}) to derive the current with respect to the gauge potential $A_t^{(0)}$. Near the horizon, the
current becomes anomalous while it is still conserved outside the horizon. Thus, if we split the region
outside the horizon into two parts: $[r_+, r_+ +\varepsilon)$ and $[r_+ +\varepsilon, +\infty)$, and
introduce two step functions $\Theta(r) = \Theta(r -r_+ -\varepsilon)$ and $H(r) = 1 -\Theta(r)$, then
we can write the total current in terms of the conserved one $J_{(O)}^{(0)\mu}$ and the anomalous one
$J_{(H)}^{(0)\mu}$ as
\be
 J^{(0)\mu} = J_{(O)}^{(0)\mu}\Theta(r) +J_{(H)}^{(0)\mu}H(r) \, ,
 \label{gfd}
\ee
where $J_{(O)}^{(0) \mu}$ and $J_{(H)}^{(0)\mu}$ satisfy $\nabla_\mu J_{(O)}^{(0)\mu} = 0$ and Eq.
(\ref{Gpa}), respectively. Both equations can be solved as
\bea
 \sqrt{-g}J_{(O)}^{(0)r} &=& c_O^{(0)} \, , \nn  \\
 \sqrt{-g}J_{(H)}^{(0)r} &=& c_H^{(0)} +\frac{e}{2\pi}
  \big[\mathcal{A}_t -\mathcal{A}_t(r_+)\big] \, ,
\eea
where the charge flux $c_O^{(0)}$ and $c_H^{(0)}$ are two integration constants. Substituting Eq.
(\ref{gfd}) into Eq. (\ref{Gpa}), the Ward identity becomes
\be
 \p_r\big[\sqrt{-g}J^{(0)r}\big] = \p_r\big(\frac{e}{2\pi}\mathcal{A}_tH\big)
 +\big\{\sqrt{-g}\big[J_{(O)}^{(0)r} -J_{(H)}^{(0)r}\big]
 +\frac{e}{2\pi}\mathcal{A}_t\big\}\delta(r -r_+ -\varepsilon) \, .
\ee
In order to make the current anomaly free under the gauge transformation, the first term must be
cancelled by the classically irrelevant ingoing modes and the second term should vanish at the
horizon, which reads
\be
 c_O^{(0)} = c_H^{(0)} -\frac{e}{2\pi} \mathcal{A}_t(r_+) \, .
\ee
Imposing the boundary condition that the covariant current vanishes at the horizon, which means
$c_H^{(0)} = 0$, then the charge flux with respect to the gauge potential $A_t^{(0)}$ is given by
\be
c_O^{(0)} = -\frac{e}{2\pi}\mathcal{A}_t(r_+) \, .
\ee
Similarly, the current corresponding to the gauge potential $A_t^{(1)}$ is
\be
 c_O^{(1)} = -\frac{m}{2\pi}\mathcal{A}_t(r_+) \, .
\ee
From Eq. (\ref{Gpa}), one can see that $J^{(0)r}$ and $J^{(1)r}$ are not independent for there
exists the relation $\frac{1}{e}J^{(0)r} = \frac{1}{m}J^{(1)r} = \mathcal{J}^{r}$ between
them, where $\mathcal{J}^{\mu}$ satisfies the covariant gauge anomaly equation
\be
 \nabla_\mu \mathcal{J}^{\mu} = \frac{-1}{4\pi\sqrt{-g}}\epsilon^{\alpha \beta}
 \mathcal{F}_{\alpha \beta} \, .
\ee
By analogy, the charge flux at spatial infinity can be obtained as
\be
 \mathcal{J}^{r}(r\to \infty) = c_O = -\frac{1}{2\pi}\mathcal{A}_t(r_+)
 = -\frac{1}{2\pi}\Big(\frac{eQ}{\Xi r_+} +\frac{ma}{\Xi l^2}\Big) \, .
\ee

With the expression of the charge flux in hand, we now pay our attention to the energy momentum flux.
Near the horizon, if we omit the quantum effect of the ingoing modes, the effective field theory will
exhibit a gravitational anomaly. For the right-handed fields, the ($1 + 1$)-dimensional covariant
gravitational anomaly is expressed as \cite{AWBZBK}
\be
 \nabla_{\mu}T_{~\nu}^{\mu} = \frac{-1}{96\pi}\sqrt{-g}\epsilon_{\mu\nu}\p^{\mu}R
 = \frac{1}{\sqrt{-g}}\p_{\mu}N_{~\nu}^{\mu} = \bar A_\nu \, .
\ee
In the case of a background spacetime with the effective metric (\ref{em1}), the anomaly is
timelike ($\bar A_r = 0$), and
\bea
 \bar A_t = \frac{1}{\sqrt{-g}}\p_{r}N_{~t}^{r} \, , \qquad
 N_{~t}^{r} = \frac{1}{96\pi}\big(fh^{\prime\prime}
  -\frac{fh^{\prime 2}}{h} +\frac{1}{2}h^{\prime}f^{\prime}\big) \, .
\eea
Now we take the effect from the charged field into account. The energy momentum outside the horizon
satisfies the Lorentz force law
\be
 \nabla_{\mu}T_{(O)\nu}^{\mu} = \mathcal{F}_{\mu\nu}\mathcal{J}_{(O)}^{\mu} \, ,
 \label{gat1}
\ee
while the energy momentum near the horizon obeys the anomalous Ward identity after adding the
gravitational anomaly,
\be
 \nabla_{\mu}T_{(H)\nu}^{\mu} = \mathcal{F}_{\mu\nu}\mathcal{J}_{(H)}^{\mu} +\bar A_\nu \, .
 \label{gat2}
\ee
Writing the total energy momentum tensor as
\be
 T_{~\nu}^{\mu} = T_{(O)\nu}^{\mu}\Theta(r) +T_{(H)\nu}^{\mu}H(r) \, ,
 \label{emf}
\ee
we solve both equations for the $\nu = t$ component and get
\bea
 \sqrt{-g}T_{(O)t}^{r} &=& a_O +c_O\mathcal{A}_t \, , \label{aoah1} \\
 \sqrt{-g}T_{(H)t}^{r} &=& a_H +\Big(c_O\mathcal{A}_t +\frac{1}{4\pi}\mathcal{A}_t^2
 +N_{~t}^{r}\Big)\Big|_{r_+}^{r} \, , \label{aoah2}
\eea
where $a_O$ and $a_H$ are two constants. Similar to the case of the charge current, we insert Eq.
(\ref{emf}) into (\ref{gat2}) and only consider the $\nu = t$ component as before, then we arrive
at
\bea
 \sqrt{-g}\nabla_{\mu}T_{~t}^{\mu} &=& c_O\p_r\mathcal{A}_t
  +\p_r\Big[\big(\frac{1}{4\pi}\mathcal{A}_t^2 +N_{~t}^{r}\big)H\Big] \nn \\
 && +\Big[\sqrt{-g}\big(T_{(O)t}^{r} -T_{(H)t}^{r}\big) +\frac{1}{4\pi}\mathcal{A}_t^2
  +N_{~t}^{r}\Big]\delta(r -r_+ -\varepsilon) \, .
\eea
The first term is the classical effect of the background electric field for constant current flow.
The second term should be cancelled by the quantum effect of the classically irrelevant ingoing modes.
In order to keep energy momentum tensor anomaly free, the third term must vanish at the horizon,
which yields
\be
 a_O = a_H +\frac{1}{4\pi}\mathcal{A}_t^2(r_+) -N_{~t}^{r}(r_+)
 = a_H +\frac{1}{4\pi}\mathcal{A}_t^2(r_+) +\frac{1}{192\pi}h^{\prime}(r_+)f^{\prime}(r_+) \, .
\ee
This equation is not enough to fix $a_O$ completely. It is necessary to impose the boundary condition
that the covariant energy momentum tensor vanishes at horizon, i.e., $a_H = 0$. Therefore, the Hawking
flux of the total energy momentum tensor at spatial infinity is
\be
 T_{~t}^{r}(r\to \infty) = a_O = \frac{1}{4\pi}\mathcal{A}_t^2(r_+) +\frac{\kappa^2}{48\pi} \, ,
\ee
from which the Hawking temperature can be determined as
\be
 T = \frac{\kappa}{2\pi} = \frac{1}{4\pi}\sqrt{h^{\prime}(r_+)f^{\prime}(r_+)}
 = \frac{f^{\prime}(r_+)}{4\pi\Xi}\, .
\ee

\section{Hawking radiation of higher dimensional
black strings through covariant anomalies}\label{hdbs}

In this section, we shall consider the case of higher dimensional black strings in AdS spacetimes.
In $d$-dimensions, there are, in general, $n \equiv [(d-1)/2]$ rotation parameters. Parameterized
by the mass $M$, the charge $Q$, and the rotation parameters $a_i$, the metric and gauge potential
for the higher dimensional black string solution take the following forms \cite{Awad}
\bea
 ds^2 &=& -\Big[\Xi^2f(r) +\big(1-\Xi^2\big)\frac{r^2}{l^2}\Big]dt^2
  -2\Big[\frac{r^2}{l^2} -f(r)\Big]\Xi a_idtd\phi^i \nn \\
 &&\quad +\Big\{\Big[\frac{r^2}{l^2} -f(r)\Big]a_ia_j +r^2\delta_{ij}\Big\}d\phi^id\phi^j
  +\frac{dr^2}{f(r)} +r^2\delta_{ab}dz^adz^b \, , \nn \\
 A &=& \frac{Q}{r^{d-3}}\sqrt{\frac{d-2}{2d-6}}\big(\Xi dt -a_id\phi^i\big) \, ,
 \label{hdbsm}
\eea
where
\be
 f(r) = \frac{r^2}{l^2} -\frac{2M}{r^{d-3}} +\frac{Q^2}{r^{2d-6}} \, ,
 \qquad \Xi = \Big(1 +\sum_{i=1}^n \frac{a_i^2}{l^2}\Big)^{1/2} \, .
\ee
In the above equations, indices $i, j = 1, ..., n$, and $a, b = 1, ..., d-2-n$. $\delta_{ab}dz^adz^b$
is the Euclidean metric on $(d-2-n)$-dimensional sub-manifold with the volume $V_{d-2-n}$, which can
be set to unit for simplicity. The coordinates $\phi^i$ assume the values $0 \leq \phi^i < 2\pi$.
The metric (\ref{hdbsm}) can be regarded as black strings or black branes with toroidal, cylindrical
or planar horizons, relying on its global identifications.

As before, we first study thermodynamics of black strings in higher dimensions. The horizon of the
black strings is determined by the equation $f(r) = 0$, namely, the largest positive root of $f(r_+) = 0$.
The angular velocities and electrostatic potential on the horizon are
\be
 \Omega_i = \frac{a_i}{\Xi l^2} \, , \qquad
 \Phi = \sqrt{\frac{d-2}{2d-6}}\frac{Q}{\Xi r_+^{d-3}} \, ,
\ee
while the Bekenstein-Hawking entropy and the surface gravity of the horizon are given by
\be
 S = \frac{\mathcal{V}_{d-2}}{4}\Xi r_+^{d-2} \, , \qquad
 \kappa = \frac{f^{\prime}(r_+)}{2\Xi} = \frac{(d-1)r_+^{2d-4}
 -(d-3)Q^2l^2}{2\Xi l^2r_+^{2d-5}} \, ,
\ee
where $\mathcal{V}_{d-2} = (2\pi)^nV_{d-2-n}$ is the volume of the hypersurface with constant
$t$ and $r$.

It is of no difficulty to check that the mass, the angular momenta and the electric charge of
black strings in higher dimensions
\bea
&& \mathcal{M} = \frac{\mathcal{V}_{d-2}}{8\pi}\big[(d-1)\Xi^2 -1\big]M, \qquad
  \mathcal{J}^i = \frac{(d-1)\mathcal{V}_{d-2}}{8\pi}\Xi Ma_i \, , \nn \\
&& \mathcal{Q} = \sqrt{2(d-2)(d-3)}\frac{\mathcal{V}_{d-2}}{8\pi}\Xi Q \, ,
\eea
obey the differential and integral Bekenstein-Smarr mass formulae
\cite{WWXD}
\bea
 &&\delta\mathcal{M} = T\delta S +\Omega_i\delta\mathcal{J}^i
  +\Phi\delta\mathcal{Q} +\Theta\delta l\, , \\
 &&\frac{d-3}{d-2}\mathcal{M} = TS +\Omega_i\mathcal{J}^i +\frac{d-3}{d-2}\Phi\mathcal{Q}
  +\frac{1}{d-2}\Theta l \, .
\eea
In order to close the first law of thermodynamics, we have introduced the generalized force
\be
 \Theta = -\frac{(d-2)r_+^{d-1} +(d-1)Ml^2(\Xi^2-1)}{8\pi l^3}\mathcal{V}_{d-2} \, ,
\ee
which is conjugate to a variable cosmological constant $l$. Remarkably, the above relations
provide a strong support for the definitions of the mass, the angular momenta and the electric
charge for the black strings.

Now we turn to perform a dimension reduction of the metric (\ref{hdbsm}). As usual, we will
consider the action of a massless scalar field near the horizon
\bea
 S[\varphi] &=& \frac{1}{2}\int dtdrd^n\phi d^{d-2-n}z ~r^{d-2}\varphi
 \Big\{-\frac{\Xi^2r^2 +l^2(1 -\Xi^2)f(r)}{r^2f(r)} \nn \\
 &&\quad \times \Big[\mathcal{D}_t +\frac{r^2/l^2 -f(r)}{\Xi^2r^2 +l^2(1 -\Xi^2)f(r)}\Xi
  a_i\mathcal{D}_{\phi^i}\Big]^2 +\frac{1}{r^{d-2}}\p_r\big[r^{d-2}f(r)\p_r\big] \nn \\
 &&\quad +\frac{1}{r^2}\Big[\delta^{ij} -\frac{r^2/l^2 -f(r)}{\Xi^2r^2 +l^2(1-\Xi^2)f(r)}a_ia_j\Big]
  \mathcal{D}_{\phi^i}\mathcal{D}_{\phi^j} +\frac{1}{r^2}\delta^{ab}\p_{z^a}\p_{z^b} \Big\}\varphi \nn \\
 &\simeq& \frac{1}{2}\sum_{m} \int dtdr ~r^{d-3}\sqrt{\Xi^2r^2 +l^2(1 -\Xi^2)f(r)}
  \varphi_m \Big\{ \nn \\
 &&\quad -\frac{\sqrt{\Xi^2r^2 +l^2(1 -\Xi^2)f(r)}}{rf(r)}\Big[\p_t
  +\frac{ieQ\Xi r^{5-d}}{\Xi^2r^2 +l^2(1 -\Xi^2)f(r)}\sqrt{\frac{d-2}{2d-6}} \nn \\
 &&\quad +\frac{im^ia_i\Xi[r^2/l^2 -f(r)]}{\Xi^2r^2 +l^2(1 -\Xi^2)f(r)}\Big]^2
  +\frac{rf(r)}{\sqrt{\Xi^2r^2 +l^2(1 -\Xi^2)f(r)}}\p_r^2 \Big\}\varphi_m \, . \qquad~~
\eea
In the process of dimension reduction, we have implemented a partial wave decomposition $\varphi =
\sum_{m}\varphi_m(t, r)\exp(im^i\phi^i +ik_{z^a}z^a)$, in which $m = \{m^i\}$. The above procedure
demonstrates that the physics near the horizon can be effectively depicted by an infinite collection
of massless scalar fields in terms of the following background metric and $n + 1$ gauge potentials
\bea
 ds^2 &=& - F(r)dt^2 +\frac{dr^2}{F(r)} \, , \qquad\quad
  F(r) = \frac{rf(r)}{\sqrt{\Xi^2r^2 +l^2(1 -\Xi^2)f(r)}} \, , \label{hdem} \\
 \mathcal{A}_t &=& eA_t^{(0)} +m^iA_t^{(i)} \nn \\
 &=& \frac{eQ\Xi r^{5-d}}{\Xi^2r^2 +l^2(1 -\Xi^2)f(r)}\sqrt{\frac{d-2}{2d-6}}
  +\frac{m^ia_i\Xi[r^2/l^2 -f(r)]}{\Xi^2r^2 +l^2(1 -\Xi^2)f(r)} \, ,
  \label{hdgp}
\eea
together with the dilaton $\Psi = r^{d-3}\sqrt{\Xi^2r^2 +l^2(1 -\Xi^2)f(r)}$, which makes no
contribution to the anomalies. In the effective gauge potential (\ref{hdgp}), $A_t^{(i)}$ are
the induced $U(1)$ gauge fields originated from the $n$ independent axial isometries. Each induced
gauge field has an azimuthal quantum number serving as the corresponding charge for each partial
mode. So there are, in total, the $n + 1$ gauge potentials yielding $n + 1$ gauge currents
$J^{(0)r}$ and $J^{(i)r}$ which satisfy Eq. (\ref{Gpa}).

It is a position to carry out the anomaly analysis for the higher dimensional black strings. The
analysis is completely in parallel with that in the preceding section, but it is simpler than ever,
since $\sqrt{-g} = 1$ for the effective metric (\ref{hdem}) in the case considered now. Therefore,
following the procedure in the last section, it is easy to obtain the gauge currents
\be
 c_O^{(0)} = -\frac{e}{2\pi}\mathcal{A}_t(r_+) \, , \qquad
 c_O^{(i)} = -\frac{m^i}{2\pi}\mathcal{A}_t(r_+) \, .
\ee
With the aid of the definition $\mathcal{J}^{r} = \frac{1}{e}J^{(0)r} = \frac{1}{m^i}J^{(i)r}$,
the charge flux can be computed as
\be
 \mathcal{J}^{r}(r\to \infty) = c_O  = -\frac{1}{2\pi}\mathcal{A}_t(r_+)
 = -\frac{1}{2\pi}\big(e\Phi +m^i\Omega_i\big) \, .
\ee
Analogously, the energy flux of energy momentum tensor at spatial infinity can be derived as
\be
 T_{~t}^{r}(r\to \infty) = a_O = \frac{1}{4\pi}\mathcal{A}_t^2(r_+) +\frac{\kappa^2}{48\pi} \, ,
\ee
from which the Hawking temperature is deduced as
\be
 T = \frac{\kappa}{2\pi} = \frac{F^{\prime}(r_+)}{4\pi} = \frac{f^{\prime}(r_+)}{4\pi\Xi} \, .
\ee

\section{Summary}

In this Letter, we have studied thermodynamical and thermal properties of the ($3 + 1$)-dimensional
rotating charged black strings and their higher dimensional extensions in AdS spacetimes. Firstly,
we have generalized the first law of thermodynamics to the case of a variable cosmological constant.
Then we applied the covariant anomaly cancellation method to derive the charge and energy momentum
fluxes from the horizon of these black strings, whose horizons are non-spherical but can be toroidal,
cylindrical or planar, according to their global identifications. In both cases of the black strings,
after performing a partial wave decomposition and a dimension reduction, the scalar field theory near
the horizon of the original spacetime can be effectively described by the ($1 + 1$)-dimensional chiral
theory, in which each rotation symmetry induces an effective $U(1)$ gauge potential with the azimuthal
quantum number acting as the charge of this gauge symmetry. Therefore, the ($1 + 1$)-dimensional field
theory can be regarded as that of the charged particles interacting with the effective $U(1)$ background
gauge field.

Our results imply that the method of anomaly cancellation can succeed in deriving the Hawking fluxes
from arbitrary dimensional AdS spacetimes with non-spherical horizons. The derivation of Hawking
radiation only relies on the covariant anomalies for gauge current and energy momentum tensor, together
with the regular requirements of the covariant charge current and energy momentum flux at the horizon.
Besides, our anomaly analysis presented in this article is very general since we do not restrict ourselves
to the simple case where the determinant of the reduced ($1 + 1$)-dimensional effective metric is equal
to one.

\section*{Acknowledgments}

S.Q.-Wu was partially supported by the Natural Science Foundation of China under Grant No. 10675051.

\section*{Appendix: Non-uniqueness of the reduced two-dimensional effective metrics}
\renewcommand{\theequation}{A.\arabic{equation}}
\setcounter{equation}{0}

In this appendix, we take the four-dimensional case as an example to show that the resulted
two-dimensional effective and physically equivalent metrics are non-unique in the process of
dimensional reduction, and they can differ by a regular, conformal factor.

The near-horizon limit of the action in Eq. (\ref{effact}) suggests that it can be effectively
replaced by the following action
\be
 S[\varphi] \equiv \frac{1}{2}\sum_m \int dtdr ~\Psi\varphi_m
  \Big[-\sqrt{\frac{g_{rr}}{-g_{tt}}}\big(\p_t +i\mathcal{A}_t\big)^2
  +\sqrt{\frac{-g_{tt}}{g_{rr}}}\p_r^2\Big]\varphi_m
  \label{apea}
\ee
in the two-dimensional spacetime with an effective metric $ds^2 = g_{tt}dt^2 +g_{rr}dr^2$.
Compared Eqs. (\ref{effact}) with (\ref{apea}), we get
\be
\Psi\sqrt{\frac{g_{rr}}{-g_{tt}}} = \frac{\Xi^2r^2 -a^2f(r)}{f(r)} \, , \qquad
\Psi\sqrt{\frac{-g_{tt}}{g_{rr}}} = r^2f(r) \, .
\ee
From these equations, we can solve the dilaton factor as $\Psi = r\sqrt{\Xi^2r^2 -a^2f(r)}$,
and obtain
\be
\sqrt{\frac{-g_{tt}}{g_{rr}}} = \frac{rf(r)}{\sqrt{\Xi^2r^2 -a^2f(r)}} \, .
\ee
With the help of the determinant $\sqrt{-g} = \sqrt{-g_{tt}g_{rr}} = G(r)$, where $G(r)$ is
an arbitrary function regular at the horizon, the metric components can be obtained as follows
\be
g_{tt} = -\frac{rf(r)}{\sqrt{\Xi^2r^2 -a^2f(r)}}G(r) \, , \qquad
g_{rr} = \frac{\sqrt{\Xi^2r^2 -a^2f(r)}}{rf(r)}G(r) \, .
\ee
Thus the reduced two-dimensional effective metric can be written as a conformal form as
\be
 ds^2 = G(r)\Big[-\frac{rf(r)}{\sqrt{\Xi^2r^2 -a^2f(r)}}dt^2
 +\frac{\sqrt{\Xi^2r^2 -a^2f(r)}}{rf(r)}dr^2\Big] \, ,
 \label{crm}
\ee
with $\sqrt{-g} = G(r)$ acting as the conformal factor. Therefore, the resulted two-dimensional
effective metric (\ref{crm}) is non-unique in the process of dimensional reduction.

To ensure the physical equivalence of the reduced two-dimensional metric (\ref{crm}) with the
original four-dimensional one (\ref{bs4d}), the zeros and signs of the $tt$ and $rr$ components
of the metric (\ref{bs4d}) in the dragging coordinate system should not be changed in the process
of dimensional reduction. This means that $G(r_+) \not= 0$. On the other hand, since the surface
gravity $\kappa$, $\sqrt{-g}J^r$ and $\sqrt{-g}T_{~t}^r$ are all conformal invariant physical
quantities in two dimensions, one must demand $G(r\to \infty) = 1$ to obtain consistent physical
results of the Hawking fluxes under the conformal transformation. There are a lot of regular
functions $\sqrt{-g} = G(r)$ that posses these two properties. One of the alternatives is $\sqrt{-g}
= r/{\sqrt{\Xi^2r^2 -a^2f(r)}}$ in Section \ref{fdbs}. Another simple and natural choice is
$\sqrt{-g} = 1$, as done in the higher dimensional case in Section \ref{hdbs}. In these two
cases, the metrics differ by a regular conformal factor $\sqrt{-g} = r/{\sqrt{\Xi^2r^2 -a^2f(r)}}$,
both of them give the consistent results in the anomaly analysis.

\end{document}